\documentclass[aps,prb,showpacs,showkeys,nofootinbib,preprint,floatfix,groupedaddress]{revtex4}
\usepackage{latexsym} \usepackage{amsmath} \usepackage{dcolumn}
\usepackage{bm} \usepackage{graphicx} \usepackage{epsfig}
\usepackage{float}
\usepackage[]{graphicx}
\usepackage[]{wrapfig}
\usepackage[]{color}
\begin{document}

%\documentclass[preprint,showpacs,preprintnumbers,amsmath,amssymb]{revtex4}
%\title {Implementation of Hubbard Model in Real-space Green's function Theory}
\title{GW+$U$ real-space Green's function calculations of x-ray spectra}
\author{Towfiq Ahmed} \affiliation{Dept.\ of Physics, Univ.\ of
Washington Seattle, WA 98195}

\author{J. J. Kas } \affiliation{Dept.\ of Physics, Univ.\ of
Washington Seattle, WA 98195}

\author{J. J. Rehr} \affiliation{Dept.\ of Physics, Univ.\ of
Washington Seattle, WA 98195}

%\authorinfo{Send correspondence to Towfiq Ahmed\\   E-mail: atowfiq@u.washington.edu}
%>>>> when using amstex, you need to use @@ instead of @

\date{\today}

%%%%%%%%%%%%%%%%%%%%%%%%%%%%%%%%%%%%%%%%%%%%%%%%%%%%%%%%%%%%%

%%%%%%%%%%%%%%%%%%%%%%%%%%%%%%%%%%%%%%%%%%%%%%%%%%%%%%%%%%%%%

\begin{abstract}
The Hubbard model 
is implemented in real-space Green's function calculations of x-ray spectra
using an effective self-energy
adapted from the LSDA+$U$ method of Anisimov {\it et al}. This self-energy
consists of an energy-dependent many-pole approximation to the GW self-energy
with an additive correction due to on-site Coulomb repulsion among the
partially filled localized-electron states.
This leads to a GW+$U$ approach which provides an efficient procedure to
account for local correlation effects on x-ray spectra.
%and excited state electronic structure
%of materials such as transition
%metal oxides (TMOs) and high T$_c$ cuprates.
 Results are presented for the spin and angular momentum projected
density of states of MnO, NiO, and La$_{2-x}$Sr$_x$CuO$_4$ (LSCO), for
the K-edge x-ray spectra of O atoms in MnO and NiO, and the
unoccupied electronic states and O K-edge spectra of undoped LSCO.
The method is found to yield reasonable agreement with 
experiment.
\end{abstract}


\maketitle
%%%%%%%%%%%%%%%%%%%%%%%%%%%%%%%%%%%%%%%%%%%%%%%%%%%%%%%%%%%%%
\section{INTRODUCTION}

Density functional theory (DFT) together with quasi-particle corrections
has been remarkably successful in describing the electronic
structure and band-gaps of weakly interacting systems. For such
systems, these quasi-particle corrections are often well described in
terms of Hedin's 
GW self-energy\cite{OnidaRMP2000,hedin69} (where $G$ refers to the
one-particle Green's function and $W$ the screened Coulomb interaction).
 On the other hand, the GW
approach is generally inadequate to describe the band gap and other
electronic properties in materials with well localized
$3d$ or $4f$ electrons.\cite{Anis91} 
On the other hand, the strong local Coulomb interactions in these systems
can often be approximated using a Hubbard-model,\cite{Anis91} in which
the on-site electron-electron repulsion is represented  by
spin- and orbital-occupancy dependent ``Hubbard parameters" $U$
and $J$. Combining the local spin density approximation (LSDA) of DFT
with the Hubbard model leads to the LSDA+$U$ method.
%which has been effective in correcting the insulating gap1
%in many correlated-electron systems.
In practice, the Hubbard correction is added to the original
Kohn-Sham LSDA Hamiltonian while an approximate mean-field term is
subtracted to avoid double-counting.\cite{Anisimov97} 
In order to calculate the excited state properties and x-ray spectra of
correlated systems, it is desirable to go beyond LSDA+$U$ and incorporate
energy dependent self-energy effects in terms of Hedin GW self-energy.
This is the approach adopted here, which we refer to as the
GW+$U$ method.  A related approach has been proposed 
by Jiang {\it et al}, where an accurate GW self-energy is calculated
starting with LSDA+$U$. Besides providing an improved screening
model, their approach correctly predicts the band-gap in several $d$ and
$f$ electron systems.\cite{Patrick2010,Patrick2009}
In another prescription, Bansil {\it et al} developed a self-consistent GW+$U$ 
scheme based on the tight-binding approximation and
a single-band Hubbard model.\cite{tanmoy,bobparamagnon}
Their method is found to qualitatively explain several pre-edge 
spectral features in high T$_c$ cuprates.\cite{susmita,Towfiq2010}   
%Following Refs.~\onlinecite{Anis91}
%,\onlinecite{Patrick2010} and \onlinecite{Patrick2009},
%a simplified form of this correction can be derived, such that its application
%becomes straightforward as a correction to the GW self-energy.
%This leads to an extension of the LDA+$U$ method for excited states
%and x-ray spectra,
%which we refer to as GW+$U$. 
%as previously discussed by 
%other authors,\cite{Anisimov97,Tanmoy2010,Patrick2010} 
Here we add Hubbard correction terms to an approximate many-pole GW
self-energy\cite{kas07} in a single-step approach,
although formally such corrections could be 
incorporated within a self-consistent GW framework.\cite{Anisimov97} 
The implementation of our approach within the real-space Green's function
(RSGF) formalism simplifies the calculation compared to conventional
LSDA/GW+$U$ methods, and  is one of the primary goals of this paper.
With the aid of this extension, we investigate the
effects of correlated $d$-electron states on 
the angular momentum projected density of states ($l$DOS), and the
excited state spectra including
x-ray absorption spectra (XAS) and x-ray emission spectra (XES) of
a number of materials.  As in other codes, e.g., WIEN2K\cite{Hugel2000} and
SPRKKR,\cite{Ebert2003} the Hubbard parameters are here
treated as fitting parameters. 

Our RSGF/GW+$U$ method is tested on several 
$d$-electron systems including MnO, NiO
and the undoped high T$_c$ cuprate La$_{2-x}$Sr$_x$CuO$_4$ (LSCO).
In these materials, the electronic structure and band gaps are strongly
influenced by $U$, the charge-transfer energy $\Delta$, and the one-electron
band-width $W$.  Related calculations for MnO and NiO have
also been carried out using a combined GW/LDA+$U$ approach by
Jiang {\it et al}.\cite{Patrick2010} Treatments of Ti oxide compounds
using LDA+$U$ within the multiple scattering formalism
have also been carried out by Kr\"uger.\cite{Kruger_2009} 
%\textcolor{red}{
%A single-band Hubbard model based paramagnetic self-energy has also been 
%developed by 
%Bansil {\it et al} within the self-consistent quasiparticle-GW (QP-GW) 
%scheme.\cite{tanmoy, bobparamagnon}
%This self-energy has been shown to capture key features of strong electronic 
%correlations in various cuprate spectroscopies including
%XAS,\cite{Towfiq2010} ARPES,\cite{susmita} optical,\cite{tanmoy} and neutron
%scattering,\cite{Arun3} in good agrement with experiments in both electron
%and hole doped High T$_c$ cuprates. 
%}
%Two commonly used experiments to measure the
%insulating gap between occupied and unoccupied states include
%valence band photo-emission (PES) and inverse photo-emission (BIS),
%but these are highly surface sensitive.\cite{Kurmaev2008} 
We find that our approach yields reasonable agreement with bulk-sensitive
probes such as XES and XAS which are used to measure band gaps
between occupied and unoccupied states.\cite{Kurmaev2008}

\section{THEORY AND METHODOLOGY}

 In this Section we describe our implementation of the GW+$U$ method
as an extension of the real-space Green's function (RSGF)
multiple-scattering formalism.\cite{rehrpccp,rehr2009} Our implementation
follows the strategy used in the FEFF9 code
and thus permits calculations of both electronic structure and
x-ray spectra that can account for local atomic-correlation
effects. Hartree atomic units ($e=\hbar=m=1$) are implicit unless
otherwise specified. 

\subsection{RSGF Method}

We begin with a brief outline of the RSGF formalism used in this work.
In this approach
physical quantities of interest are expressed in terms of the local
quasi-particle Green's function $G({\bf r, r'},E)$. For example,
the physical quantity measured in XAS for photons
of polarization ${\bf \hat \epsilon}$ and energy $\omega$ %$\hbar\omega=E-E_c$
is the x-ray absorption coefficient $\mu(\omega)$,
\begin{equation}
\mu(\omega) \propto \ - \frac{2}{\pi} {\rm Im}\, \left< \phi_c|\,{\hat {\bf
{\epsilon}}}\cdot{\bf {r}} \, G({\bf{r}} , {\bf{r'}} , \omega + E_{c})\, {\hat
{\bf{\epsilon}}}\cdot{\bf {r'}}|\phi_c \right> ,
\end{equation}
 where $E_c$ is the core electron energy
and $\left|\phi_c\right>$ is the core state wave function.
%and
The FEFF9 code also calculates closely related quantities such
as the spin and angular momentum projected density of states
($l$DOS) $\rho^{(n)}_{l\sigma}(E)$ at site $n$,
\begin{equation}
\rho_{l\sigma}^{(n)}(E) = \ - \frac{1}{\pi} {\rm Im}\,
\sum_m \int_{0}^{R_n} G_{L,L}^{\sigma,\sigma}(r,r, E) \, r^2 \, dr,
\end{equation}
where $R_n$ is the Norman radius around the n$^{th}$ atom,\cite{feff84ref} which is 
analogous to the
Wigner-Seitz radius of neutral spheres, and
the factor $2$ accounts for spin
degeneracy. The coefficients $G^{\sigma,\sigma'}_{L,L'}$ characterize
the expansion of
the Green's function $G({\bf {r}},{\bf {r'}},E)$ in spherical harmonics,
%% The angular dependence of the
%% real space Green's function can also be expressed using spherical harmonics $Y_L$ as
\begin{equation}
  G({\bf {r}},{\bf {r'}},E) = \sum_{L,L',\sigma} Y_L(\hat{\bf r}) \, G_{L,L'}^{\sigma,\sigma}(r,r',E) \, Y^{*}_{L'}(\hat{\bf r}'),
\end{equation}
where $L=(l,m)$
denotes both orbital and azimuthal quantum numbers.
In these formulae, the quasi-particle Green's function for an
excited electron at energy $E$ is given formally (matrix-indices suppressed)
by
\begin{equation}
G(E)=\left[{E - H -\Sigma(E)}\right]^{-1},
\end{equation}
where $H$ is the independent-particle Hamiltonian
\begin{equation}
H={\frac{p^{2}}{2}}+V,
\end{equation}
and $V$ is the Hartree-potential.
%and exchange-correlation $V_{xc}$ potentials, while
%$\Sigma(E)$ denotes the energy-dependent, one-electron self-energy.
For convenience in our calculations, the Hamiltonian is re-expressed
in terms of a Kohn-Sham Hamiltonian $H^{KS} = H + V_{xc}$ where
$V_{xc}$ is a ground state exchange-correlation\cite{Barth_Hedin} 
functional, and the self-energy is replaced by a modified self-energy
$\Sigma(E)-V_{xc}$ which is set to zero at the Fermi-energy $E=E_F$.
In this work we use the von Barth-Hedin LSDA
functional $V_{xc}[n({\bf r}),m({\bf r})]$,\cite{Barth_Hedin}
where $n({\bf r}) = n_{\uparrow}+n_{\downarrow}$ is the total 
electron density and $m({\bf r}) = n_{\uparrow}-n_{\downarrow}$ 
is the spin polarization density.
In practice, it is useful to decompose the total Green's
function ${\it {G}}(E)$ as
\begin{equation}
G(E) = G^{c}(E) + G^{sc}(E),
\end{equation}
where $G^{c}(E)$ is the contribution from the central
(absorbing) atom and $G^{sc}(E)$ is the scattering part.
Full multiple scattering (FMS) calculations can be carried out by
matrix inversion, i.e., with $G = [1-G^0 T]^{-1} G^0$, where $G^0$ is the
bare propagator and $T$ is the scattering T-matrix, which are represented
in an angular-momentum and site basis: $G^0= G^0_{nL,n'L'}(E)[1-\delta_{n,n'}]$
and $T=t^{\sigma}_{nL} \delta_{l,l'}\delta_{m,m'}\delta_{n,n'}\delta_{\sigma,\sigma'}$. 
Finally,
$t^{\sigma}_{nL}$ is the single site scattering matrix, which is related to the
single site phase shifts, i.e.,  
$$t^{\sigma}_{nL} = \exp({i\delta^{\sigma}_{nL}})\sin(\delta^{\sigma}_{nL}).$$
%where $L=(l,m)$
%denotes both orbital and azimuthal quantum numbers.
Within the spherical muffin-tin approximation, $G^{c}(E)$ can be
expanded in terms of  
%of the radial functions 
% where $R_L$ and $N_L$ are
the regular $R_L({\bf {r}},E)$ and irregular $H_L({\bf{r}},E)$
solutions of the single site %Dirac 
Schr{\" o}dinger equation.\cite{feff82ref} 
%for local spherical potential approximation
%to $V$.
%% The angular dependence of the
%% real space Green's function can also be expressed using spherical harmonics $Y_L$ as
%% \begin{equation}
%% G({\bf {r}},{\bf {r'}},E) = \sum_{L,L'} Y_L(\hat{\bf r}) \, G_{L,L'}(r,r',E) \, Y^{*}_{L'}(\hat{\bf r'}).
%% \end{equation}
In the FEFF code a typical calculation of the electronic
structure (ground or excited state) starts with a self-consistent calculation of
the electron density and Kohn-Sham potentials.\cite{feff84ref}
%The potential $V^{crys}$ of the system is first calculated self-consistently
%using multiple scattering formalism for the ground state.\cite{feff84ref}
Once the self-consistent potential is obtained, the Green's function
is constructed and used to calculate XAS and other quantities of
interest. Of particular interest in this paper is the local
spin-dependent density matrix for the $n$-th site
%$n^{\sigma \sigma'}_{mm'}$ 
%is obtained using the relation
\begin{equation}
\label{eq:densitymatrix}
n^{\sigma\sigma^{'}}_{nlm,nlm^{'}} = \ - \frac{1}{\pi} \int^{E_{F}} dE
\int_{cell} \! {\rm Im}\, G^{\sigma\sigma^{'}}_{nlm,nlm^{'}} ({\bf {r}},{\bf{r}},E) \,d^{3}r, 
\end{equation}
where the $n$ denotes the cell defined by the Norman sphere centered
about the $n^{th}$ atom,
${\bf r}, {\bf r'}$ are relative to the center of the cell $R_{n}$, and
$\sigma$ is the spin-index, and we explicitly designate the azimuthal
quantum numbers $m$  and $m'$.
%where $G^{\sigma}_{ilm,ilm^{'}} = \left< ilm\sigma|(E- {\hat {H}})^{-1}|ilm^{'}
%\sigma\right>$ 
%are the elements of the Green function matrix in this localized 
%representation. 
For a more detailed description of the multiple scattering RSGF method see
Refs. [\onlinecite{rehr00,feff82ref}].  

\subsection{GW+$U$ Self-energy}

Quasi-particle effects are key to an accurate treatment of excited state
spectra,\cite{rehrpccp} and hence we need a good approximation for the
electron self-energy. Current approximations for the self-energy typically
begin with Hedin's $GW$ approximation (GWA),\cite{hedin69}
which is formally given by
\begin{equation}
 % \Sigma = iGW\Gamma \approx iGW,
  \Sigma = i G W
\end{equation}
where $G$ is the one electron Green's function, $W = \epsilon^{-1} v,$
is the screened-Coulomb interaction, and $v$ the bare-Coulomb interaction.
%and $\Gamma$ is
%the vertex correction which is replaced by the unit operator in the GWA.
% state what GW is
The FEFF9 code uses several approximations for the self-energy 
%$\Sigma^{GW}=i G W$,
%where $G$ is the Green's function and $W$ is the screened Coulomb interaction.
with the aim of providing efficient calculations of the
energy dependent shift and broadening of spectral features.  The default,
which is appropriate at high energies, is the Hedin-Lundqvist
plasmon-pole model,\cite{hedin69,plasmon3}
based on the electron gas and a single-pole approximation to the dielectric
function.  An extension which improves the self energy at low energies is
a many-pole model, where the
the dielectric function is represented as a weighted sum of poles 
% This many-pole model dielectric function is then
matched to calculations of the loss function in the long
wavelength limit.\cite{kas07}
%%Here we note that the $\Sigma^{GW}(E)$ in our RSGF formalism has been adopted
%%from the LDA approximation of plasmon-pole model,\cite{hedin69,plasmon3} and an 
%%extension 
%%of this with a many-pole model.\cite{kas07}  
Although these models significantly improve
quasi-particle calculations of unoccupied states, they do not
necessarily obtain accurate band-gap corrections.
%A simple method of
%obtaining an appropriate gap correction is to use the model of
%Ref.~\onlinecite{wangpickett84}, with the gap energy set by the LDA+$U$
%calculation. 
%%A self-energy is then modified 
%%by adding a state dependent Hubbard correction term, as
%%in Eq.\ (8), 
In our implementation of GW+$U$, 
an energy, spin and orbital dependent total potential 
%an additional energy dependent 
%self-energy 
is constructed that incorporates the GW plasmon-pole or many-pole
self energy $\Sigma(E)$ and the Hubbard correction $V^U_{lm}$,
with parameters chosen to obtain the correct gap.
Although such a construction can be done using
self-consistent methods,\cite{georges_2004} here we use only a
single-step calculation. Thus we define our total potential as
\begin{equation}
V({\bf r},E) = V_{\sigma}^{LSDA}({\bf r})+\Sigma^{GW}(E)+V^U_{lm\sigma}.
\end{equation}  
%The Hubbard contribution can then be incorporated into our Green's
%function approach, by constructing a self-energy of the form
%where
%\begin{align}
%\Sigma^{U}_{lm\sigma}(E)=&V^{U}_{lm\uparrow}\Theta(E_{b}-E)+V^{U}_{lm\downarrow}
%\Theta(E-E_{t}) \nonumber \\
%&+i\gamma\Theta(E-E_{b})\Theta(E_{t}- E).
%\end{align}
%Here $E_{b} = E_{F} - V_{lm\uparrow}$, $E_{t} = E_{F}+V_{lm\downarrow}$,
%$\gamma$ is a large constant which
%forces the density of states in the gap to zero, and
%$\Theta(x)=1$ or $0$ for $x>0$ or $x<0$ respectively. Note that this
%formula essentially mimics the effects of the shifted
%eigenstates in wave-function based implementations of the
%Hubbard model.
The orbital and spin-dependent Hubbard contribution to the
potential 
is calculated as described in the next section.
%The above approach is essentially equivalent to wave-function methods which
%incorporate the Hubbard correction in terms of a state dependent potential.
%and where the orbital occupancies are
%obtained from RSGF calculations of the
%density matrix using Eq.\ (10).

\subsection{LSDA+$U$ formalism }

Our construction of $V^{U}_{lm\sigma}(E)$ is adapted from the LSDA+$U$
approach of Anisimov {\it et al}.\cite{Anisimov97} In their approach
one starts with the total
energy functional of the system and adds a Hubbard correction 
to account for the Coulomb interaction between localized, strongly
correlated electrons. It is generally assumed\cite{Albers2009} that a
similar mean-field term should exist in LSDA or other DFT approaches
which must be subtracted from the 
energy functional to avoid double counting,
\begin{eqnarray}
E^{U}[n^{\sigma}(\vec{r}),{\bf n^{\sigma}}]
&=& E^{LSDA}[n^{\sigma}(\vec{r})]   \\ \nonumber 
&+& E^{U}[{\bf n^{\sigma}}]-E_{dc}[{\bf n^{\sigma}}],
\end{eqnarray}
where $n^{\sigma}(\vec{r})$ is the charge density, ${\bf n^{\sigma}}$ the
density matrix, $E^{U}$ the Hubbard interaction, and E$_{dc}$ the double 
counting term.  The Hubbard term depends on the density matrix
$n^{\sigma \sigma'}_{ilm,ilm'}$, 
and on-site Coulomb interactions between the localized electrons. 
%such as the
%the direct term $U_{mm'm''m'''}=\langle m,m'|V_{ee}|m'',m'''\rangle $ and 
%the exchange term $J_{mm'm'''m''}=\langle m,m'|V_{ee}|m''',m''\rangle $.

As discussed by Albers {\it et al}.,\cite{Albers2009} an {\it ab initio}
determination of the Hubbard parameters is not straightforward,
since they are sensitive to screening of the Coulomb interaction.
Hence, the Hubbard terms are often regarded as fitting parameters while
the density matrix is calculated from first-principles to construct
the Hubbard potential. This is the approach adopted here.
While it may be possible to go beyond this
parametrization and calculate the Hubbard terms using approaches such as
constrained-LDA\cite{Gunnarsson1989,Solovyev2005}, or
constrained-RPA,\cite{Aryasetiwan2006,Cococcioni2005},
such estimates are beyond the scope of this paper.
For systems where the localized electrons are atomic-like, the
density matrix can be  approximated\cite{Anisimov93} as
\begin{equation}
n^{\sigma}_{mm'}=n^{\sigma}_m \delta_{mm'}.
\end{equation}
%%With this spherical approximation, the off-diagonal terms of the
%%density matrix $n^{\sigma}_{mm'}$ are ignored. 
This spherical approximation is reasonable for many systems including
TMOs, and good agreement for the band gap is
found when the non-sphericity of d-d interaction as well as the off-diagonal 
terms of $n_{mm'}$ are ignored.\cite{Anis91}
%% so that
%% \begin{align}
%% U_{mm'm''m'''} \rightarrow U_{mm'}=  
%%     \langle mm'|V_{ee}|mm'\rangle \rightarrow U , \nonumber \\  
%% J_{mm'm''m'''} \rightarrow J_{mm'}= 
%%    \langle mm'|V_{ee}|m'm\rangle \rightarrow J  .  
%% \end{align}
%[comment: Is the above about ignoring the ``non-sphericity'' of U and
%  J, or about ignoring the off diagonal terms of $n_{mm'}$?]
With these approximations, the number of parameters is reduced to only
two, namely $U$ and $J$ representing the screened direct and exchange
intra-atomic Coulomb interactions, respectively.

The total energy functional can then be written as
\begin{eqnarray}
E&=&E^{LSDA} + \frac{1}{2} \sum_{m,m',\sigma} U(n^{\sigma}_{m}-n^o)
(n^{-\sigma}_{m'}-n^o) \nonumber \\ 
&+& \frac{1}{2} \sum_{m,m' \neq m,\sigma} (U-J)(n^{\sigma}_{m}-n^o)
(n^{\sigma}_{m'}-n^o).
\end{eqnarray}       
Here the double counting term $E_{dc}$ is represented by $n^o$ where 
$n^o=n_d/10$ and $n_d=\sum_{m \sigma}n^{\sigma}_m$.
Using $V(\vec{r})=\delta E/\delta n_{\sigma}(\vec{r})$, a simplified expression
for the total LSDA+$U$ potential is finally obtained,\cite{Anisimov93}
i.e.,
\begin{equation}
V^{LSDA+U}(\vec r) = V^{LSDA}(\vec r) + V^{U}_{lm\sigma}, 
\end{equation}
where
\begin{equation}
V^{U}_{lm\sigma} = 
    U\sum_{m'} ( n_{lm^{'}}^{-\sigma}-n_{o}) 
 + (U-J) \sum_{m' \neq m} (n_{lm^{'}}^{\sigma}-n_{o}) .
\end{equation}
%This potential correction is then added to
%our GW self-energy below.

%% The potential $V^{crys}$ of the system is first calculated self-consistently
%% using multiple scattering formalism for the ground state.\cite{feff84ref}
%% Once the self-consistent potential is obtained, the Green's function
%% is constructed and the density matrix for $i$-th site  
%% %$n^{\sigma \sigma'}_{mm'}$ 
%% is obtained using the relation
%% \begin{equation}
%% n^{\sigma\sigma^{'}}_{ilm,ilm^{'}} = \ - \frac{1}{\pi} \int^{E_{F}} dE
%% \int_{cell} \! {\rm Im}\, G^{\sigma\sigma^{'}}_{ilm,ilm^{'}} ({\bf {r}},{\bf{r}},E) \,d^{3}r, 
%% \end{equation}
%% where $G^{\sigma}_{ilm,ilm^{'}} = \left< ilm\sigma|(E- {\hat {H}})^{-1}|ilm^{'}
%% \sigma\right>$ are the elements of the Green function matrix in this localized 
%% representation. 
Within the spherical approximation, we need only consider the diagonal
elements $n^{\sigma}_{lm}$ of the 
density matrix defined in Eq.~(\ref{eq:densitymatrix}). 
In a single-step spin-dependent calculation using the von Barth-Hedin LSDA
functional, we first
obtain $n_{lm}^{\sigma}$. In this prescription, a prior knowledge of spin 
polarization of $i$-th atom $m_i = n^{\uparrow}_i - n^{\downarrow}_i$ is
required.   For Mn, Ni, and Cu we used $m$ = 5, 2, and 1 correspondingly using 
Hund's multiplicity 
rule\cite{Hunds1,Hunds2} for free atoms which is often treated as good approximations
for such systems.  

%Using a spin-independent calculation we first calculate $n_{lm}$ using
%\begin{equation}
%n_{lm}=n^{\uparrow}_{lm}+n^{\downarrow}_{lm}. 
%\end{equation}
The occupancy of the spin-up and -down states within the $d$-orbitals
are thus determined in this single-step LSDA approach. 
Our calculations of spin-orbital occupancies of Mn and Ni $d$-states using this
scheme are listed in Tables I and II. 
%\textcolor{red} {
Thus we essentially start with a spin dependent 
ground state calculation and introduce spin and orbital dependence using 
Anisimov's prescription of Hubbard model. 
This LSDA+$U$ prescription is found to provide 
good agreement between the theory and experiment for the XAS of the
TM compounds investigated here, although the self-consistent
LSDA+$U$ treatment may be more desirable in other cases.
%}  
%\cite{Hunds1,Hunds2} 
%which is a good approximation for localized atomic-like
%$d$-states in TMOs.\cite{Hunds1,Hunds2}  

Values for the $U$ and $J$ parameters are taken either from
previous work\cite{Anis91,constrained_RPA2006} or chosen to fit
the experimental band gap. 
%\textcolor{red}{
For MnO and NiO, we used $U \approx 4.5$ and $7.5$ eV and $J = 0.9$ eV, which
are reasonably close to those calculated or discussed
by other authors.\cite{constrained_RPA2006,Patrick2010}
%} 
Using Eqs.\ (9) and (14) we then correct our self-consistent
quasi-particle (QP) %crystal 
potential and obtain a new potential $V^{GW+U}({\bf{r}},E)$ given by
\begin{equation}
V^{GW+U}_{\sigma}({\bf {r}},E)= V^{LSDA}_{\sigma}({\bf {r}})+\Sigma^{GW}(E)+
V^U_{lm\sigma}(E) .
\end{equation}

Then using the 
%the potential in Eq.\ (8) derived from the
GW+$U$ Hamiltonian above, the wave functions
$R_L({\bf {r}},E)$ and $H_L({\bf{r}},E)$ 
are recalculated as solutions of the Schr{\" o}dinger equation inside 
the muffin-tin spheres with our Hubbard modified potential.
The orbital dependent phase shifts $\delta^{\sigma}_{lm}(E)$ are
obtained by matching to the free solutions (spherical Bessel
functions) at the muffin-tin, and the   
scattering $t$-matrices are found.  
\begin{equation}
t^{\sigma}_{lm}=\exp({i\delta^{\sigma}_{lm}})\,\sin(\delta^{\sigma}_{lm}).
\end{equation}
%where, the $t$-matrix for the $n$-th site is defined formally
%by the Dyson equation
%$t_n=[1-v_n G^0]^{-1}v_n,$
%where $G^0$ is the free propagator. %or two site Green's function.
Finally the multiple-scattering equations are resolved with these
$t$-matrices yielding the
the total Green's function $G=G^c+G^{sc}$, which now includes 
the Hubbard-$U$ correction.
{\color{red}  }
With the addition of the state dependent Hubbard correction, the
potential of Eq.~(15) can correctly account for the well known
discontinuity\cite{Perdew82,Anisimov93} in exact 
DFT exchange-correlation potentials. However, such a term is absent
from the conventional LDA and GGA approaches,
rendering them incapable of including such
band-gap corrections.

\section{RESULTS AND DISCUSSION}

\subsection{Transition Metal Oxides}  
Transition metal oxides (TMOs) such as MnO and NiO are considered to be
prototypes of strongly correlated Mott type insulators with localized and partially
filled $d$-electrons at the metal sites. These TMOs have NaCl like 
crystal structures, (Cubic $O^5_h$ symmetry, and f$_{m3m}$ space group).
Below their
respective N\`eel temperatures, they all exhibit a rhombohedral distortion due 
to anti-ferromagnetic (AF)
ordering, which is also known as exchange anisotropy.\cite{berkowitz99}
We examined the effects of such crystal distortions but they had
negligible influence on the spectral features of interest here.
In the following subsections 
we present results for the total and angular momentum projected DOS of
MnO and NiO for a few values of $U$. The exchange parameter $J$ is typically
much smaller than $U$ and variations were found to be small over the
transition metals; thus we used $J=0.9$ eV for all cases.\cite{Anis91}
For both compounds, the O K-edge 
XAS and XES are also calculated and compared with experimental results.
%Different spectral features are then discussed
%%and compared with experiment.\cite{Kurmaev2008} 

\subsubsection{{\bf\rm MnO}}

%Below the N\`eel temperature $T_N=116 K$, MnO crystals
%exhibit a rhombohedral distortion due to AF ordering.\cite{Morosin70} 
%The experimental and theoretical description of this phenomenon  has
%been discussed elsewhere.
%\begin{table}[ht]  
%\caption{Mn $d$-state occupancy}
%\centering
%\begin{tabular*}{0.35\textwidth}{@{\extracolsep{\fill}} c  r  c  c  c  r  r}
%\hline
%{\it l} &  m  &  $n_{lm}$  &  $n_{lm}^{\uparrow}$ & $n_{lm}^{\downarrow}$ & $V_{lm}^{\uparrow}$(eV) & $V_{lm}^{\downarrow}$(eV) \\ [0.5ex] 
%\hline\hline 
%2 & $\pm2$ & 0.76 & 0.76 & 0.00 & -1.38 & 1.65\\
%2 & $\pm1$ & 1.16 & 1.00 & 0.16 & -2.32 & 1.02\\ 
%2 & 0 & 0.31 & 0.31 & 0.0 & 0.40 & 1.65\\[1ex] 
%\hline
%\end{tabular*}
%\end{table}
In order to compare with room temperature experiment,\cite{Kurmaev2008} we
have taken an undistorted MnO crystal with $a = b = c = 4.4316$ \AA\ and 
${\it \alpha} = {\it \beta} = 90.624^{\circ}$.\cite{Morosin70}  
In this paper, we do not consider periodic magnetic effects;
however, the single site moments are implicitly taken into account
in our GW+$U$ implementation. 
\begin{figure} [h] 
 \includegraphics[scale=0.34,angle=-90]{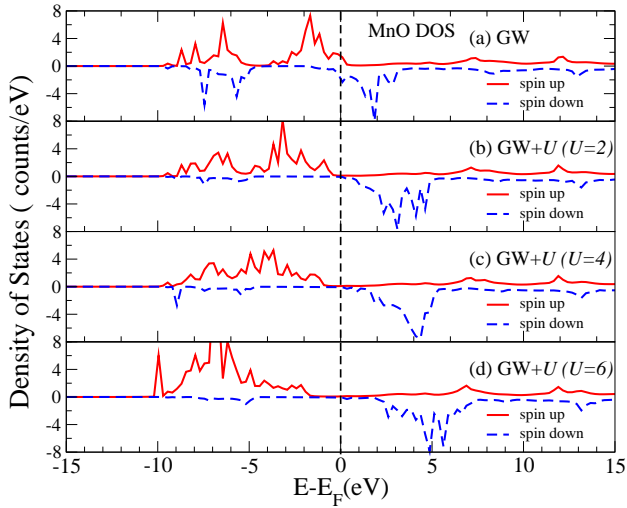}
  \caption
   {
(Color online) $U$ dependence on total DOS of MnO with spin up (solid red) 
and spin down (dashed blue) for different values of $U$: 
(a) GW ($U$=0), (b) $U=2.0$ eV, (c) $U=4.0$ eV and (d) $U=6.0$ eV;
Vertical dashed line is at the Fermi energy. 
}
\end{figure}
\begin{figure}  
 \includegraphics[scale=0.34,angle=-90]{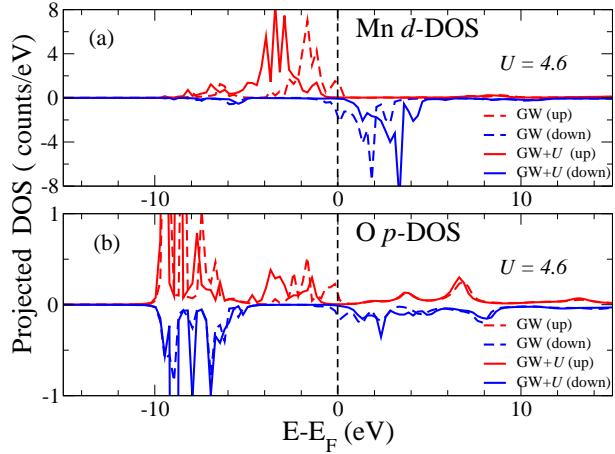}
  \caption
   {(Color online) angular momentum projected $l$-DOS for Mn and O in MnO
with GW($U=0.0$) 
and GW+U ($U=4.6$ eV); Spin up an down DOS are above and below the horizontal axis 
correspondingly: 
(a) Mn $d$-DOS GW(dashed) and GW+$U$ (solid). (b) similar color and line style
correspondence for O $p$-DOS; Vertical dashed line is at the Fermi energy. 
}
\end{figure}
In our FMS calculations for MnO, we used a cluster of 250 atoms,
which was adequate to converge the spectrum, and a
smaller cluster of 60 atoms for the self-consistent muffin-tin potentials.
%% Our FMS calculations for MnO with an extension of FEFF9 used
%% a finite cluster with 250 atoms, which is adequate to converge the 
%% spectrum.
For this system we calculated  the
%the density of states and the
O K edge XES and XAS  
%The muffin-tin potential
%was calculated self-consistently with a smaller cluster of 60 atoms. 
%Once self-consistency was achieved for the crystal 
%potential, 
and the spin and angular 
momentum projected DOS about the Mn and O sites with and without GW+$U$  
corrections. Fig.\ 1 shows a comparison of our calculated total ground state 
spin-resolved DOS of MnO to 
that calculated with different values of $U$.
\begin{figure}  
 \includegraphics[scale=0.34,angle=-90]{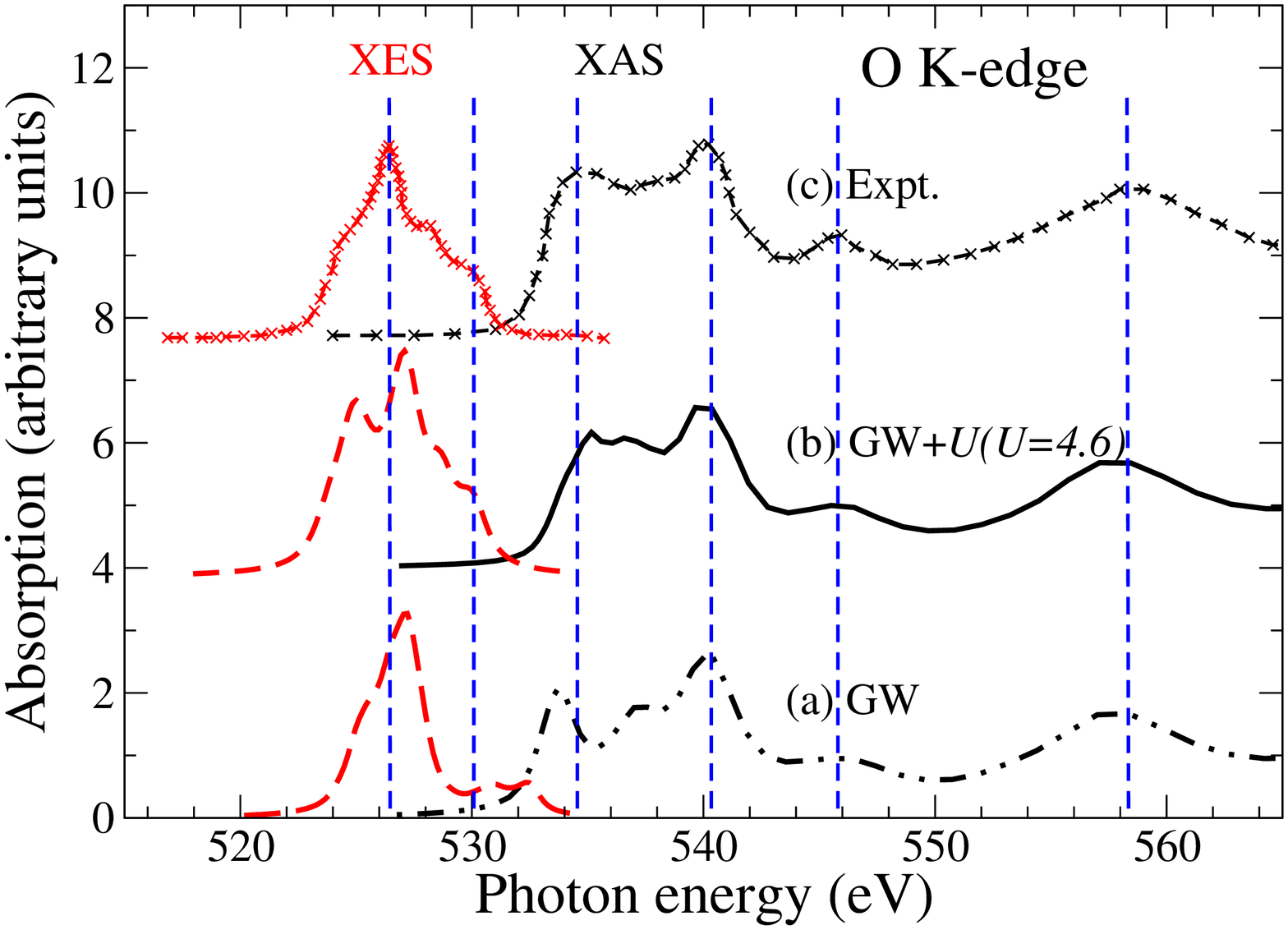}
  \caption
   {(Color online)  
O K-edge XAS (black) and XES (red) in MnO. (a) QP
FEFF calculation using GW plasmon-pole 
self-energy, (b) GW+$U$ ($U=4.6$), and 
(c) experiment.\cite{Kurmaev2008}
The vertical dashed lines are a guide to the eye. 
}
\end{figure}
While a calculation with a GW self-energy underestimates an
insulating gap (dashed blue line in Fig.\ 1), a gap close to that
observed in experiment is obtained using the Hubbard correction $U = 4.6$ eV.
When applied to Mn $d$-states, an upper Hubbard
band appears at about +1.5 eV, as seen in Fig.\ 2(a). The
O $p$-states (Fig.~2(b)) near E$_F$ are strongly hybridized with
Mn $d$-states (Fig.\ 2(a)); thus a gap is also seen in the O
$p$-DOS. However, the O $p$-states around 6-8 eV 
only hybridize with Mn $s$-$p$-states (not shown) and are not affected 
by the Hubbard correction. In Table (I) we  present the spin-orbital 
occupancies of the localized Mn $d$-states and the corresponding Hubbard 
correction for $U=4.6$ and $J=0.9$ eV.  
\begin{table}[ht]  
\caption{Mn $d$-state parameters ($U=4.6$; $J=0.9$ eV)}
\centering
\begin{tabular*}{0.40\textwidth}{@{\extracolsep{\fill}} c  r  c  c  c  c  c}
\hline
{\it l} &  m  &  $n_{lm}$  &  $n_{lm}^{\uparrow}$ & $n_{lm}^{\downarrow}$ & $V_{lm}^{\uparrow}$(eV) & $V_{lm}^{\downarrow}$(eV) \\ [0.5ex] 
\hline\hline \\[-0.2ex] 
2 & 0 & 1.02 & 0.90 & 0.12 & -1.82 & 1.61\\
2 & $\pm1$ & 0.97 & 0.84 & 0.13 & -1.56 & 1.55\\ 
2 & $\pm2$ & 0.95 & 0.83 & 0.12 & -1.52 & 1.63\\ [1ex] 
\hline
\end{tabular*}
\end{table}

Bulk sensitive XES and XAS for TM oxides often provide a good 
assessment of the band gap in insulators.\cite{Kurmaev2008} In Fig.\ 3 we 
compare our GW+$U$ calculation of the O K-edge XAS and XES with
experiment.\cite{Kurmaev2008} Fig.\ 3 shows the result of our 
spin resolved FMS 
calculation obtained with both Hubbard and $GW$ self-energy corrections (b) compared to results with no Hubbard correction (a) and experiment (c). 
%Comparison with
%experiment confirms the underestimation of an insulating gap in this
%QP calculation. 
%The curves labeled as (b) is 
%our RSGF XAS calculation including both the GW plasmon pole self-energy
%and our Hubbard correction. 
The XAS calculation was done in the presence of a screened core-hole
at the absorbing O atom while for XES no core-hole was included; these
approximations are consistent with the final-state- and initial-state
rules for XAS and XES respectively.  Our Hubbard corrected
self-energy blue shifts the first excitation at around 534 eV,
while the rest of the unoccupied states, including the main peak at 540 eV, 
are unchanged. In XES, the highest occupied state moves down by 3 eV which
is now on the other side of the second vertical dashed line in Fig.\ 3. 
These distinct, opposite shifts of the highest occupied and first 
unoccupied states are
due to the strong hybridization of O $p$-states with the localized Mn $d$ 
states. This can also be identified in Fig.\ 2(b) as the lower and upper 
Hubbard bands (UHB) at around -2 and 2 eV. 
\subsubsection{{\bf\rm NiO}}
%The N\`eel temperature of NiO is $T_N=523 K$, below which it exhibits exchange
%anisotropy. 
In order to compare with room-temperature experiments\cite{Kurmaev2008} 
we have accounted for the rhombohedral distortion along the [111]
direction.\cite{exchange75,exchange64} Our methods for calculating 
electronic structures of NiO are similar to those for MnO, except
for the input NiO crystal structure, where we have used a  
slightly distorted crystal with $a = b = 4.168$ \AA, $c = 4.166$ \AA, and 
$\alpha = \beta = 90.055^{\circ} , \gamma = 90.082^{\circ}$. 
%A GW calculation of the O K-edge XAS and XES in NiO again underestimates an
%insulating gap, which is found experimentally to be $\approx$ 4.0 eV.
With the Hubbard correction, the best agreement with the experimental XAS was 
again obtained with $U$ = 7.5 eV.  Fig.\ 4 and Fig.\ 5 show 
the gap opening in the spin projected total DOS of NiO for 
higher values of $U$. 
%the gap opening, and the presence of an UHB at 2.5 eV and a LHB at -1.5 eV. 
The O $p$-states in NiO are also strongly hybridized with localized Ni 
$d$-states as in MnO. The spin-orbital occupancies and
corresponding Hubbard potential for the Ni $d$-states are listed in Table II.
\begin{table}[ht]  
\caption{Ni $d$-state parameters ($U=7.5$ eV; $J=0.9$)}
\centering
\begin{tabular*}{0.40\textwidth}{@{\extracolsep{\fill}} c  r  c  c  c  c  c}
\hline
{\it l} &  {\it m}  &  $n_{lm}$  &  $n_{lm}^{\uparrow}$ & $n_{lm}^{\downarrow}$ & $V_{lm}^{\uparrow}$(eV) & $V_{lm}^{\downarrow}$(eV) \\ [0.5ex] 
\hline\hline \\ [-0.2ex] 
2 & 0 & 1.22 & 0.97 & 0.25 & -2.75 & 3.74\\
2 & $\pm1$ & 1.64 & 0.96 & 0.68 & -2.53 &  1.88\\ 
2 & $\pm2$ & 1.44 & 0.97 & 0.47 & -2.63 & 2.48\\ [1ex] 
\hline
\end{tabular*}
\end{table}

Our GW plasmon-pole calculation in Fig.\ 6(a) exhibits considerable overlap
between the O K-edge XAS and XES spectra, due to the underestimated 
insulating gap. However, the introduction of the Hubbard interaction ($U=7.5$ eV)
increases the gap, causing the pre-peaks of both the XAS
and XES to split further apart, as shown in Fig.\ 6(b). 
For comparison, we also show a WIEN2K LDA+$U$ calculation in Fig.\ 6(c)
%(Dobysheva {\it et al.})
for the O K-edge EELS in NiO.\cite{Dobysheva2004} 
\begin{figure}  
 \includegraphics[scale=0.34,angle=-90]{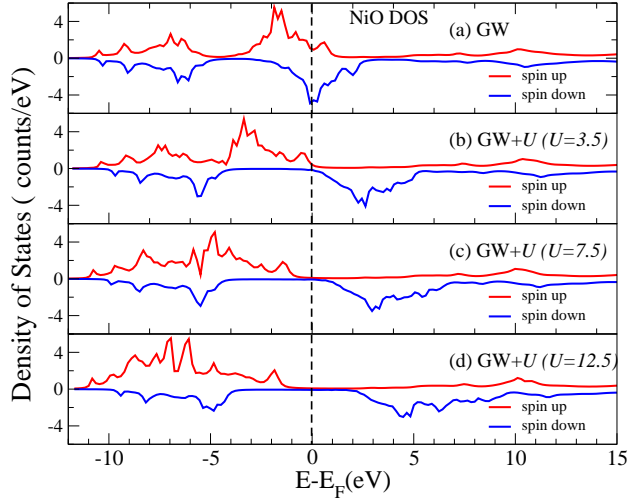}
  \caption
{
(Color online) $U$ dependence on total DOS of NiO with spin up (solid red) 
and spin down (dashed blue) for different values of $U$: 
(a) GW ($U$=0), (b) $U=3.5$ eV, (c) $U=7.5$ eV and (d) $U=12.5$ eV;
Vertical dashed line is at the Fermi energy. 
} 
%  { (Color online) Total DOS of NiO with (solid red) 
%and without $U$ (dashed black) for different values of  
%$U$: (a) $U=2.0$ eV, and (b) $U=4.0$ eV. 
%}
\end{figure}
\begin{figure}[H]  
\begin{center}
 \includegraphics[scale=0.34,angle=-90]{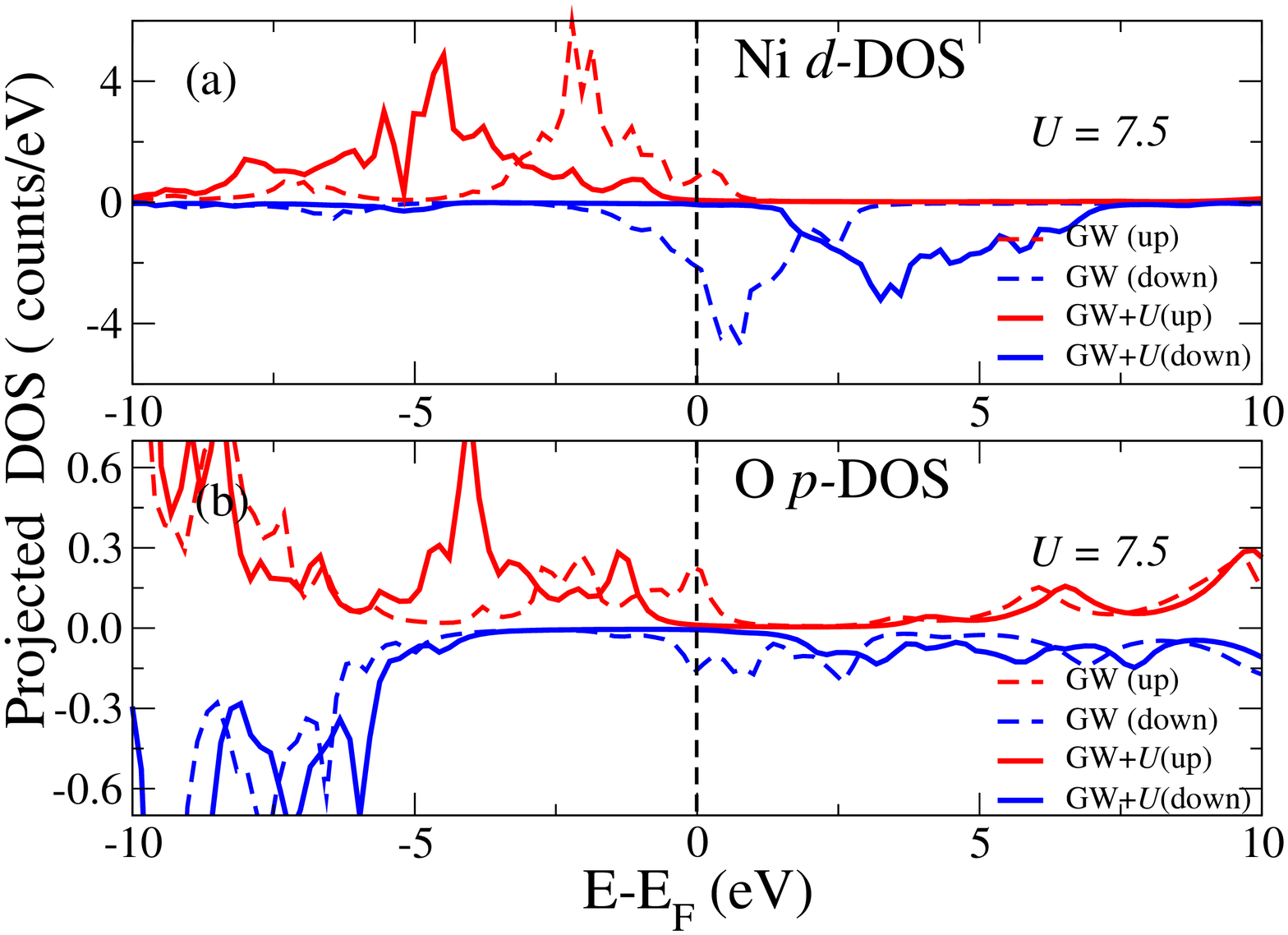}
  \caption
   {(Color online) angular momentum projected $l$-DOS for Ni and O in NiO
with GW($U=0.0$) 
and GW+U ($U=7.5$ eV); Spin up an down DOS are above and below the horizontal axis 
correspondingly: 
(a) Ni $d$-DOS GW(dashed) and GW+$U$ (solid). (b) similar color and line style
correspondence for O $p$-DOS; Vertical dashed line is at the Fermi energy. 
}
\end{center}
\end{figure}

\begin{figure}  
 \includegraphics[scale=0.34,angle=-90]{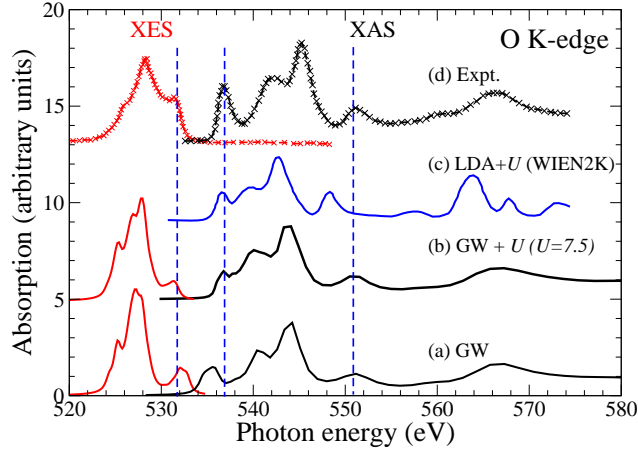}
  \caption
   {(Color online) 
NiO O K-edge XAS (black) and XES (red) experiment vs theory:
(a) FEFF GW+$U$ plasmon-pole (PP) self-energy; 
(b) FEFF GW+$U$ with Hubbard correction $U=7.5$ eV; 
(c) GW+$U$ calculation\cite{Dobysheva2004} for O 1s EELS using 
WIEN2K; and 
(d) Experiment.\cite{Kurmaev2008}
The vertical dashed lines are a guide to the eye. 
}
\end{figure}
%using the WIEN2K package.\cite{Dobysheva2004}    
Aligning the first peak of this calculation with experiment [Fig.\ 6(d)], 
we observe an underestimation of  the high energy peaks at around 544 eV.
These peaks can be attributed to O $p$-states
which are strongly hybridized with  Ni $s-$ and $p$-states.
Similar behavior has been found in NiO,\cite{Kurata93,Patrick2010} and other
TM compounds.\cite{Dobysheva2004, Kotani2008} 
% as well as calculations with plasmon-pole self-energy (not shown).
We attempted to improve these results by using a more accurate GW many-pole
self-energy\cite{kas07} for NiO,
while applying the Hubbard correction to the Ni $d$-states. 
This many-pole self-energy includes a more realistic treatment of inelastic
losses than the plasmon pole model, and yields improved agreement
with experiment, as seen from Fig.\ 6(b).
%Thus the Hubbard correction adequately accounts for the localized
%$d$-states and pre-edge 
%features of O K-edge XAS and XES,  
These results demonstrate that an accurate treatment of the
delocalized $s$-$p$-states can also be important in such systems.
Thus in order to achieve good agreement between theoretical and experimental
spectral features, a systematic consideration of excited state properties
including both localized- and delocalized states seems to be important.

\subsection{LSCO}

In recent years, understanding the doping dependence of high T$_c$ cuprates
has become an interesting challenge.  LSCO (La$_{2-x}$Sr$_x$CuO$_4$),
which is a prototype of hole-doped cuprates,
exhibits metallic and paramagnetic behavior at high 
doping,\cite{Towfiq2010} and becomes 
an AF insulator when undoped. Between these limits, the system
goes through a superconducting 
phase at about $x$ = 0.15. A good description of the electronic structure
in its insulating phase is important to understand the
exotic doping dependent phase transformations
in such systems. 

In the over-doped region with doping concentrations $x > 0.2$, LSCO 
becomes paramagnetic, and is well described by a self-energy approximation
constructed from a single band Hubbard model.\cite{Towfiq2010}
A Fermi-liquid description thus becomes more 
appropriate for such systems. As doping is reduced, 
correlation effects due to localized states become more important, and  
the implementation of Hubbard $U$ to the $d$ electrons on the Cu 
sites is seen to open a gap. 
%We fit our $U$ value to 4.0 eV to match the band gap with experiment. 
A gap correction using GW+$U$ on the partial $d$-DOS of Cu and $p$-DOS of O  
is shown in Fig.\ 7.
\begin{figure}  
 \includegraphics[scale=0.34,angle=-90]{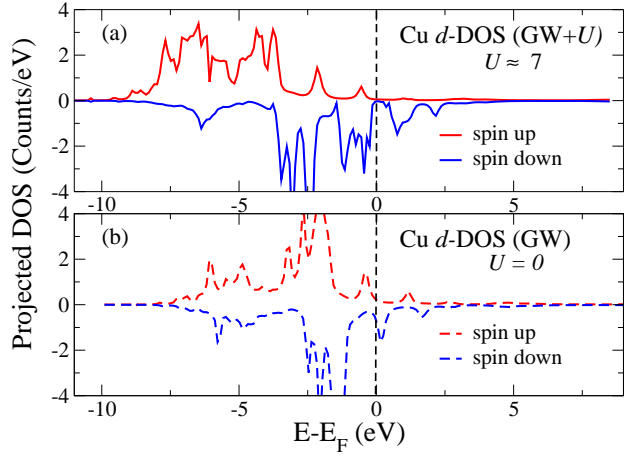}
  \caption
   {(Color online)  
(a) GW+$U$ ($U$=7.5 eV) Cu spin up (solid red) and down (dashed blue) $d$-DOS;
(b) GW ($U$=0.0) Cu spin up and down $d$-DOS. The vertical dashed line is  
at the Fermi energy. 
}   
\end{figure}
\begin{figure}  
 \includegraphics[scale=0.34,angle=-90]{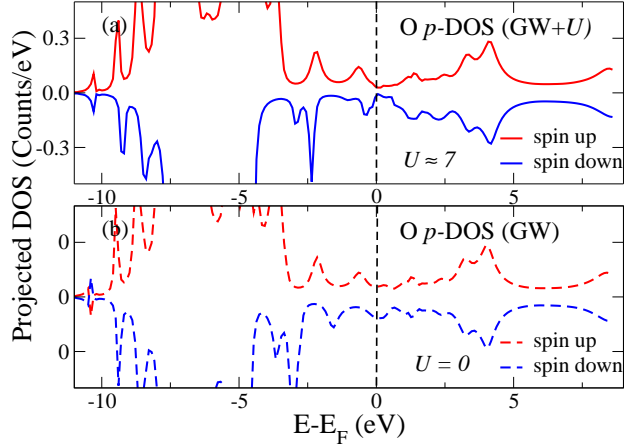}
  \caption
   {(Color online)  
(a) GW+$U$ ($U$=7.0 eV) O spin up (solid red) and down (dashed blue) $p$-DOS;
(b) GW ($U$=0.0) O spin up and down $p$-DOS. The vertical dashed line is  
at the Fermi energy. 
}   
\end{figure}
Our O K-edge XAS for GW and GW+$U$ with $U$ = 7.0 eV  are compared with
experimental results in Fig.\ 9.
Our result agrees qualitatively with the  
undoped LSCO experiment, while the over-doped LSCO system
is adequately reproduced by a GW calculation alone ($U$ = 0). 
\begin{figure}  
 \includegraphics[scale=0.34,angle=-90]{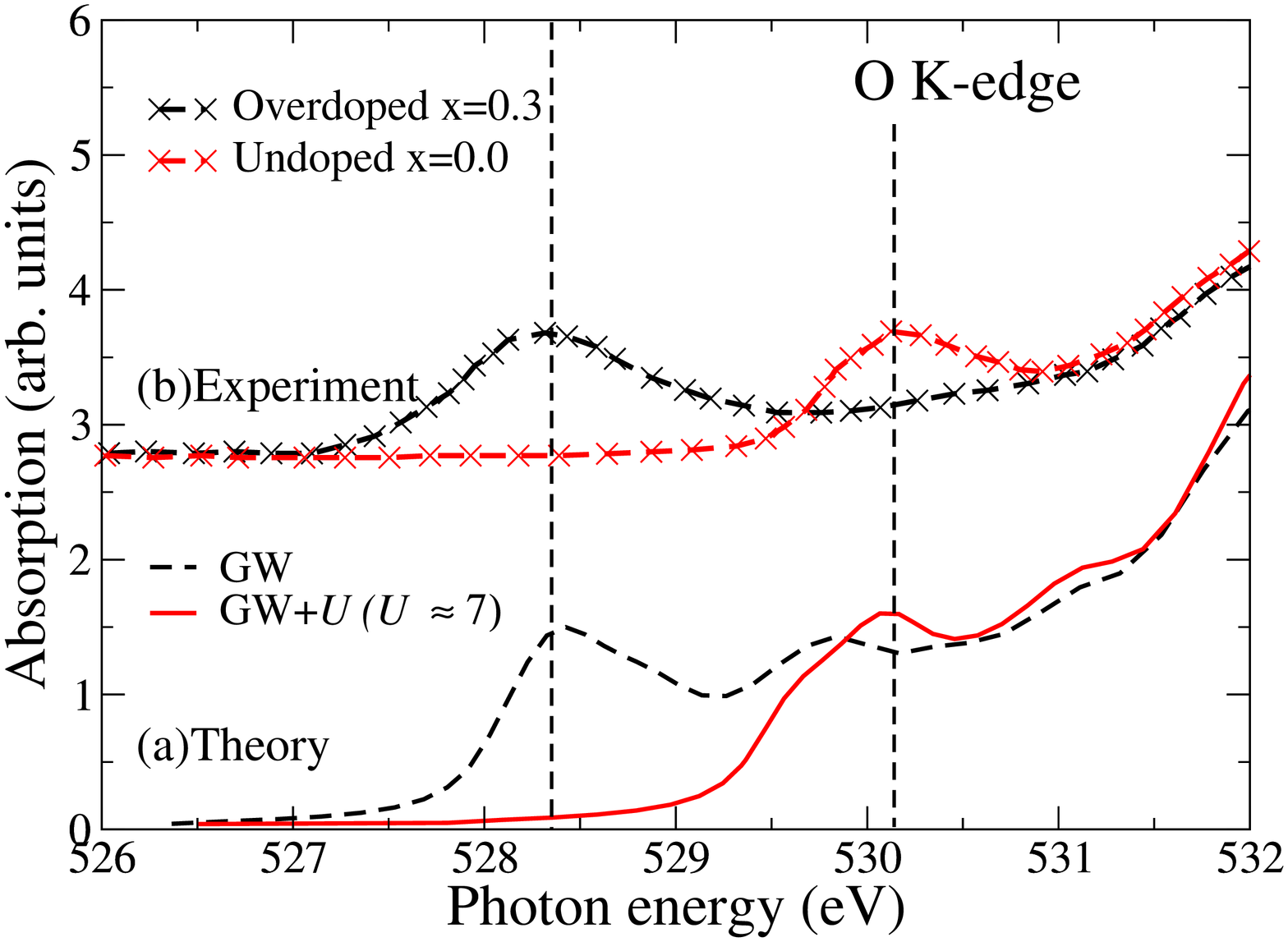}
  \caption
   {(Color online)  
O K-edge XAS for LSCO: 
(a) our GW+$U$ calculation with for $U = 7.0$ (red) and GW only (black);
(b) experimental K-edge XAS for 
undoped ($x=0.0$, red) and over-doped ($x=0.3$, black) LSCO, and  
the vertical dashed lines are a guide to the eye.  
}   
\end{figure}
This result is not surprising, since in the absence 
of the Hubbard term, the LDA does not predict a correlation gap.
As a result the system is predicted to be metallic, mimicking the 
over-doped ($x \approx$ 0.3) paramagnetic phase of La$_{1-x}$Sr$_{x}$CuO$_4$. 
A complete description of the doping dependence of spectral features
from over-doped ($x=0.3$) to undoped ($x=0.0$), requires a dynamical
self-energy correction that incorporates pseudo-gap,
superconducting, and Fermi-liquid physics.\cite{Bob2010} 
\\

\section{SUMMARY AND CONCLUSIONS}

We have implemented a Hubbard model adapted from the LSDA+$U$ method of
Anisimov {\it et al.}  as an extension of  the real-space Green's function
approach for calculations of x-ray spectra of correlated materials.
In our construction two parameters $U$ and $J$ are chosen to match
the experimental gap.
%For simple TMOs such as NiO and MnO, the Hubbard parameters $U$
These Hubbard parameters are introduced in terms of an effective self-energy
correction leading to a GW+$U$ approach which provides an efficient way
to account for local correlation effects on x-ray spectra. 
Such a
theoretical understanding of O K-edge XAS and XES is useful to 
explain key electronic features of strongly correlated systems 
%which can serve as a diagnostic of correlation effects.  
For example, in the AF
insulating phases of transition metal oxides, several important features of
the experimental profile of the O K-edge XAS and
XES, and in particular the correlation-gap, require theoretical treatments
beyond the quasi-particle approximation.
%We find that these features can be 
%accounted reasonably well within our RSGF/GW_U approach.

%nd $J$ were fitted
%o match the experimental gap. 
%In our calculations, the Hubbard interaction was only applied to the $d$-states
%of the TM sites. The resulting gap affected the O $p$-states through 
%hybridization and is reflected 
%the localized $d$-states were found to
%hybridize correctly, through multiple scattering with the rest of the 
%system including O 
%$p$-states near E$_F$. This is reflected 
%in our calculated 
%pre-edge features in the O K-edge XAS and the XES. 
Our GW+$U$ approach yields results which are in good agreement with 
experiment for the O K-edge spectrum of MnO and NiO. However, the agreement 
is only qualitative for more complex systems such as LSCO. This 
suggests the need for including a more comprehensive treatment of
superconducting and pseudo-gap physics capable of 
incorporating doping dependence in the under-doped regime of such 
systems.\cite{Tanmoy2010,Bob2010}
%The effect of doping dependence in Fig.\ 8 is only qualitative. 
%A better scenario can be captured by extending to a dynamical implementation
%of the Hubbard model in our RSGF theory as done in the
%DMFT\cite{georges_2004} approach.  
Finally we note that our current approach is limited to the quasi-particle 
approximation together with Hubbard model corrections, while many-body
effects such as satellites are neglected.
However, some of these many-body aspects can be obtained
by incorporating additional charge transfer contributions in
the Hamiltonian.\cite{Hedin_1999,Hedin_PRB99} 
%within a quasi-boson 
%formulation in our RSGF/GW+$U$ method. 

\section{ACKNOWLEDGMENT}

We thank A. Bansil and R. Markiewicz and especially P. Rinke for
stimulating suggestions.
This work is supported by the Division of Materials Science \&
Engineering, Basic Energy Sciences, US Department of Energy
Grants DE-FG03-97ER45623 and DE-FG02-07ER46352.  This research also
benefited from the collaboration supported by the Computational Materials
Science Network (CMSN) program of US DOE under grant DE-FG02-08ER46540.

%\bibliographystyle{plain}
%\bibliography{references}

\end{document}